\documentstyle[12pt]{article}
\begin{document}

\def\0{\c{s}}
\def\1{\"{o}}
\def\beq{\begin{equation}}
\def\eeq{\end{equation}}
\def\bea{\begin{eqnarray}}
\def\eea{\end{eqnarray}}
\def\ve{\vert}
\def\vel{\left|}
\def\ver{\right|}
\def\nnb{\nonumber}
\def\ga{\left(}
\def\dr{\right)}
\def\aga{\left\{}
\def\adr{\right\}}
\def\rar{\rightarrow}
\def\nnb{\nonumber}
\def\la{\langle}
\def\ra{\rangle}
\def\lla{\left<}
\def\rra{\right>}
\def\ba{\begin{array}}
\def\ea{\end{array}}
\def\ds{\displaystyle}


\title{ {\Large {\bf 
Spin Correlation Effects In a One Dimensional Electron Gas } } }

\author{\vspace{1cm}\\
{\small Murat Tas \thanks
{e-mail: tasm@metu.edu.tr}\,\, and
Mehmet Tomak \thanks
{e-mail: tomak@metu.edu.tr}} \\
{\small Physics Department, Middle East Technical University} \\
{\small 06531 Ankara, Turkey} }

\date{}

\begin{titlepage}
\maketitle
\thispagestyle{empty}

\begin{abstract}
The Singwi, Sj\1lander, Tosi, Land (SSTL) approach is generalized to 
study the spin--correlation effects in a one dimensional electron gas. 
It is shown that the SSTL approach is capable of generating results
comparable to the more widely used STLS approach.
\end{abstract}

\end{titlepage}
\section{Introduction}

The Singwi, Tosi, Land and Sj\1lander (STLS) approach\cite{STLS} is a 
powerful theoretical tool in going beyond the random phase approximation 
(RPA) in studying the short range correlation effects of an interacting 
electron gas. It was originally developed for the three dimensional (3D) 
electron gas. In the STLS approach the short range correlation effects 
are described by the local field correction $G(q)$ for the density
response function.

The STLS approach is later applied to the two 
dimensional\cite{C1}--\cite{Freitas} and the one dimensional electron
gas (1D)\cite{Campos}--\cite{Ekrem} with long range Coulomb or short 
range interactions.

The Lobo, Singwi and Tosi (LST) approach\cite{LST} was originally
developed to calculate spin correlation effects in the interacting 
3D electron gas. Although the calculated spin susceptibility is not 
in agreement with experiment, LST approach is applied to the two 
dimensional\cite{Moudgil}--\cite{Rajagopal} and one dimensional 
electron gas\cite{Bilal},\cite{Gold} problems.

In this paper, we study the validity of yet another attempt to go 
beyond RPA, the Singwi, Sj\1lander, Tosi and Land (SSTL) 
approach\cite{SSTL} which was originally developed for a 3D electron
gas. It is proposed as an improvement over the STLS approach to 
better take into account the compressibility rule. For a review 
of the STLS, SSTL and LST approaches please see reference\cite{Solid}. 
It is curious that though it is not very widely used, the SSTL 
approach is not investigated before for a low dimensional electron 
gas problem. It is therefore of interest to study the SSTL approach, 
to investigate its range of validity and to compare its results with 
the very widely used STLS approach. This is done in our earlier 
work for density correlations in a 1D electron gas\cite{Tas}.

In the present paper, we generalize the SSTL approach to study the spin 
correlation effects in a one dimensional interacting electron gas. The
organization of the paper is as follows: In section II we present the 
SSTL approach. The results for the structural properties are given in 
section III. A discussion of our results and the preformance of the 
SSTL approach is presented in section IV.

\section{Formalism}

The ground state of the electron gas at high density is paramagnetic. 
In a paramagnetic system the magnetic moments of the constituents are 
randomly distributed and their magnetic moment is averaged to zero. 
If we apply a weak external magnetic field to such a system, it will 
develop a paramagnetic spin moment. The response of the system to the 
field can be studied via its wave vector and frequency dependent 
paramagnetic susceptibility function. In noninteracting electron 
system, the Pauli paramagnetic susceptibility term is pronounced 
instead of the paramagnetic susceptibility term. On the other hand, 
in an interacting electron system we have short range Coulomb and 
exchange effects. Then, the Pauli value must be improved. In this 
study, we investigate these effects within the self--consistent 
calculation approximation (SSTL) as in the case of density--density 
correlations. Furthermore, it is assumed that our system is a Fermi 
liquid and the electrons are embedded in a uniform positive background 
so that the system is neutral. 
     
In the self--consistent calculation approximation the wave vector and
frequency dependent density and spin--density response functions, 
$\chi^d(q,\omega)$ and $\chi^s(q,\omega)$ respectively, are

\beq
\chi^d(q,\omega)=\frac{\chi_0(q,\omega)}{1-\psi^s(q)\chi_0(q,\omega)}, 
\eeq
and

\beq
\chi^s(q,\omega)=-g^2 \mu_B^2 \frac{\chi_0(q,\omega)}
{1-\psi^a(q)\chi_0(q,\omega)},
\eeq
where $\chi_0(q,\omega)$ is the free electron polarizability, 
$\psi^s$ ($\psi^a$) is the spin symmetric (spin antisymmetric) 
potential, $g$ is the Land$\acute{e}$ factor and $\mu_B$ is the 
Bohr magneton. 

The system responds to the applied magnetic field through the free
particle susceptibility modified by a local field correction. The 
static structure factor $S(q)$ and the magnetic static structure 
factor ${\ds \tilde{S}(q)}$ are related to the dynamic response 
functions by the fluctuation--dissipation theorem as follows:

\beq
S(q) = \frac{1}{\pi n} \int_0^{\infty} d\omega
~Im \left\{\chi^d(q,\omega) \right\},
\end{equation}
and 

\begin{equation}
\tilde{S}(q)=\frac{1}{\pi n g^2 \mu_B^2} \int_0^{\infty} d\omega 
~Im \left\{\chi^s(q,\omega) \right\}.
\end{equation}

In the mean field approximation, the effective potentials 
$\psi^s(q)$ and $\psi^a(q)$ are defined as

\beq
\psi^s(q)=V(q)[1-G(q)], ~~~~ \psi^a(q)=V(q)I(q),
\eeq
where $V(q)$ is the one dimensional Fourier transform of the Coulomb
potential, $G(q)$ and $I(q)$ are the static local field correction
arising from the short range Coulomb correlation and 
exchange--correlation effects for the density and spin--density
responses, respectively. In the SSTL approximation they are given
in one dimension by

\beq
G(q)=-\frac{1}{n} \int \frac{dq^{\prime}}{2\pi} \frac{q^{\prime}}{q}
\frac{V(q^{\prime})}{V(q)~\varepsilon(q)}\left[S(q-q^{\prime})-1\right],
\eeq 
and

\beq
I(q)=\frac{1}{n} \int \frac{dq^{\prime}}{(2\pi)}\frac{q^{\prime}}{q}
\frac{V(q^{\prime})}{V(q)~\varepsilon(q)}\left[\tilde{S}(q-q^{\prime})
-1\right],
\eeq
where $n$ is the one dimensional electron gas density. The Fermi wave 
vector $k_F$ is related to the linear (1D) electron density as 
${\ds n=2g_{\nu}k_F/\pi}$. The $g_{\nu}$ is the valley degeneracy and 
in this work we assume $g_{\nu}=1$, which is the case for the quantum 
wire structures $GaAl/Al_xGa_{1-x}As$. In one dimension, the 
dimensionless density parameter is defined as ${\ds r_s=\pi/4k_F 
a_B^{\star}}$, where ${\ds a_B^{\star}=\varepsilon_0/e^2m^{\star}}$ 
is the effective Bohr radius with background dielectric function 
$\varepsilon_0$ and effective electron mass $m^{\star}$. The 
$\varepsilon(q)$ is the static dielectric function which is given in 
terms of the density response function $\chi^d(q)$ as

\beq
\frac{1}{\varepsilon(q)}=1+V(q)\chi^d(q).
\eeq

In the STLS approximation, the potential under the integral sign 
in Eq. (6) and in Eq. (7) is not screened by $\varepsilon(q)$.

The random phase approximation (RPA) describes the dielectric 
properties of the electron gas very successfully at high electron
densities. In RPA the short range correlations are assumed to be 
absent, i.e, the local field correction $G(q)=0$. As the density 
of the system is lowered, the exchange and the correlation effects 
become important. The Hubbard approach (HA) was developed to 
improve the RPA by introducing a local field correction taking 
into account the exchange hole around the electron. In HA, the 
local field correction for spin correlations, $I_H(q)$, is given by

\begin{equation}
I_H(q)=-\frac{1}{2}\frac{V \left(\sqrt{q^2+k_F^2}\right)}{V(q)}.
\end{equation}

The spin symmetric and spin antisymmetric pair correlation functions 
are related to the static structure factor and the magnetic structure 
factor by a Fourier transform in any dimension respectively. In one 
dimension 

\beq
g(r)=1+\frac{1}{2} \int_0^{\infty}~dq~cos(qr) \left[S(q)-1\right],
\eeq
and

\beq
\tilde{g}(r)=\frac{1}{2} \int_0^{\infty}~dq~cos(qr)
\left[ \tilde{S}-1 \right].
\eeq
These can be written in terms of parallel spin pair correlation 
function, ${\ds g_{\uparrow\uparrow}(r)}$, and antiparallel spin 
pair correlation function ${\ds g_{\uparrow\downarrow}(r)}$ as

\beq
g(r)=\frac{1}{2}\left[g_{\uparrow\uparrow}(r)+
g_{\uparrow\downarrow}(r)\right], ~~~~ 
{\ds \tilde{g}}(r)=\frac{1}{2}\left[g_{\uparrow\uparrow}(r)-
g_{\uparrow\downarrow}(r)\right].
\eeq

The spin dependent potentials may be obtained by combining the 
$\psi^s(q)$ and $\psi^a(q)$ as

\beq
\psi_{\uparrow\uparrow}(q)=\psi^s(q)+\psi^a(q), ~~~~
\psi_{\uparrow\uparrow}(q)=\psi^s(q)-\psi^a(q).
\eeq

The static density and spin--density susceptibilities in the self 
consistent calculation approximation may be obtained easily from 
Eq. (1) and (2) as

\beq
\chi^d(q)=\frac{\rho(\varepsilon_F) k_F}{q}
~\frac{\ln \left| {\ds \frac{q-2k_F}{q+2k_F}} \right|}
{1-{\ds \frac{16 r_s k_F}{\pi^2 q}}F(q)[1-G(q)] 
\ln \left| {\ds \frac{q-2k_F}{q+2k_F}} \right|},
\eeq
and

\beq
\chi^s(q)=\frac{g^2 \mu_B^2 \rho(\varepsilon_F)k_F}{q}
~\frac{\ln \left| {\ds \frac{q+2k_F}{q-2k_F}} \right|}
{1+{\ds \frac{16 r_s k_F}{\pi^2 q}} F(q) I(q)
\ln \left| {\ds \frac{q+2k_F}{q-2k_F}} \right|}.
\eeq
where ${\ds \rho(\varepsilon_F)=2m^{\star}/\pi k_F}$ is the 
density of states at the Fermi energy in one dimensional 
electron gas. Note that the Pauli spin susceptibility is 
${\ds \chi_{Pauli}=\mu_B^2 \rho(\varepsilon_F)}$\cite{mahan}.

\section{Results and Discussion}

The results presented in this section are obtained by solving Eqs. 
(4), (7) and (8) self--consistently. In Fig. 1, the magnetic 
structure factor ${\ds \tilde{S}(q)}$ is presented for different 
$r_s$ values for a wire of effective radius ${\ds b=2a_B^{\star}}$. 
The appearance of a peak at $q=2k_F$ as the density decreases has 
also been reported in earlier studies\cite{Bilal}. It gets 
progressively difficult to achieve convergence in the 
self--consistent calculation for $r_s>1.8$. An instability sets in 
at this value of $r_s$, as also observed in earlier STLS 
studies\cite{Bilal},\cite{Gold}. This unfortunately means that we 
are unable to study different possible phases of the 1D electron 
gas as one varies the density or equivalently the $r_s$ parameter.

The magnetic structure factor ${\ds \tilde{S}(q)}$ obtained using
different approaches is shown in Fig. 2. It seems that the Hubbard
approximation has the most pronounced peak at shown values of $r_s$ 
and the wire radius $b$. The sharpness of the SSTL peak comes out to 
be the least. 

In contrast to ${\ds \tilde{S}(q)}$, the static structure factor 
$S(q)$ is easier to obtain even for larger values of $r_s$, as 
observed also in earlier work\cite{Tas}. The $S(q)$ for $r_s=2$, 
and different values of the effective wire radii are shown in Fig. 3.

The static local field corrections arising from the short range 
Coulomb correlation and exchange--correlation effects for the 
density and spin--density responses are shown in Figs. 4 and 5,
respectively.

The spin dependent pair correlation functions ${\ds 
g_{\uparrow\uparrow}(r)}$ and ${\ds g_{\uparrow\downarrow}(r)}$ are 
shown in Figs. 6 and 7, respectively, for different $r_s$ values. It 
is seen that ${\ds g_{\uparrow\uparrow}(r)}$ has very weak $r_s$ 
dependence.

The spin symmetric and anti symmetric effective potentials ${\ds 
\psi_{\uparrow\uparrow}(q)}$ and ${\ds \psi_{\uparrow\downarrow}(q)}$ 
in units of $V(q)$ are shown in Figs. 8 and 9. They are also compared 
with the STLS results. The effective potentials have rather similar 
behaviour in both cases.

The static density susceptibility function is compared with STLS result 
in Fig. 10. The static spin--density susceptibility, on the other hand, 
is compared with STLS and HA results in Fig. 11. It is seen that the 
results in all three approaches are qualitatively and quantitatively 
similar. The peak at $q=2k_F$ is less pronounced in the SSTL approach 
which seems to be its dominant character.

Collective excitations in the one dimensional electron gas can be 
studied as the poles of the density and spin--density response 
functions $\chi^{d,s}(q,\omega)$. The dispersion of the paramagnon 
peak, $\omega_p(q)$ is shown in Fig. 12 for $r_s=1.5$. For small 
$q$, $\omega_p(q)$ shows a linear behaviour, as also observed 
before in STLS work\cite{Bilal}. 

In summary, we have studied the spin correlations in a one dimensional 
electron gas, and have shown that the SSTL approach is capable of 
giving results qualitatively similar to those obtained by using the 
STLS approach. This is, of course, not enough to establish the 
performance of the SSTL approach in 1D. The question of compressibility
inconsistency should be settled before one considers the SSTL approach 
as a viable alternative to other approaches. 

\section*{Acknowledgments}
The authors thank Dr. Ceyhun Bulutay for his help in the computation 
part of this work.

\newpage

\begin{figure}
\vskip 1cm
    \includegraphics{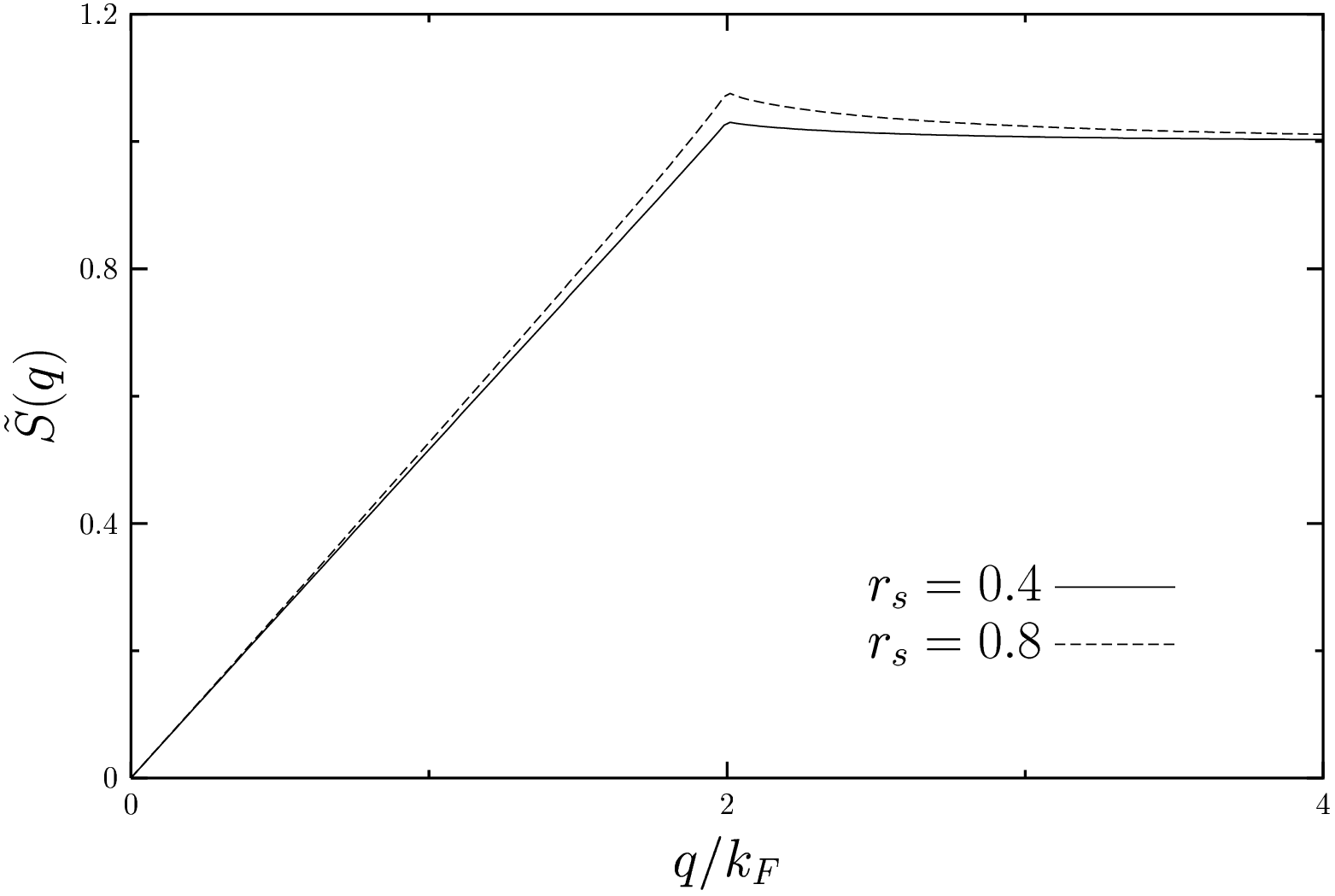}
\vskip 9.2cm
\caption{The magnetic structure factor ${\ds \tilde{S}(q)}$ for
different $r_s$ values at wire radius ${\ds b=2a_B^{\star}}$ in 
the SSTL approximation.}
\end{figure}

\begin{figure}
\vskip 1cm
    \includegraphics{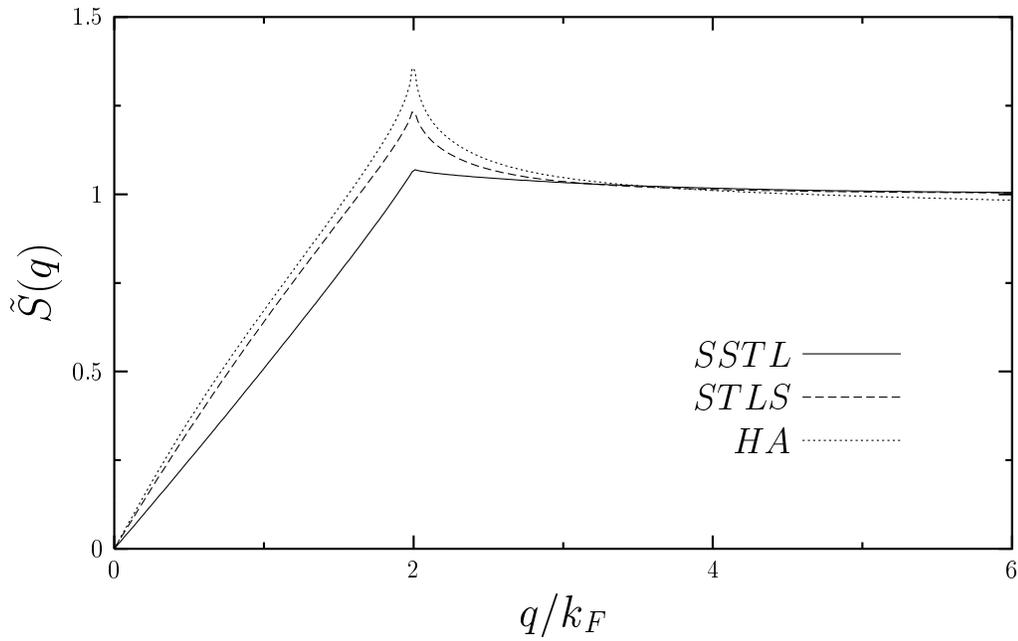}
\vskip 8.1cm
\caption{The magnetic structure factor ${\ds \tilde{S}(q)}$ at 
$r_s=1$ and ${\ds b=2a_B^{\star}}$ in different approximations.}
\end{figure}
\newpage

\begin{figure}
\vskip 1cm
    \includegraphics{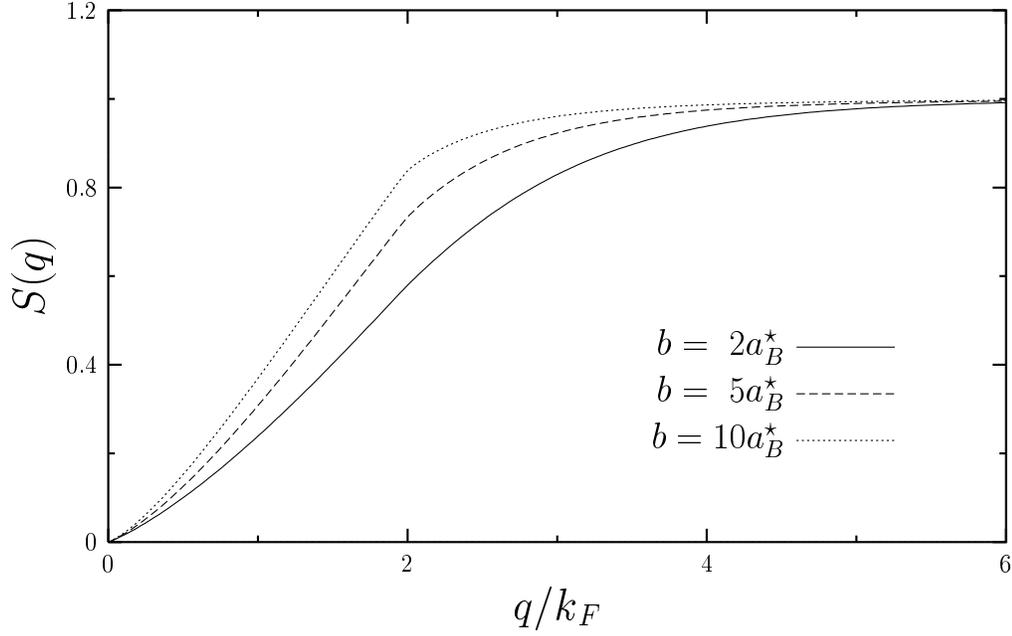}
\vskip 8.1cm
\caption{The static structure factor $S(q)$ at $r_s=2$ for different
wire radii in the SSTL approximation.} 
\end{figure}

\begin{figure}
\vskip 1cm
    \includegraphics{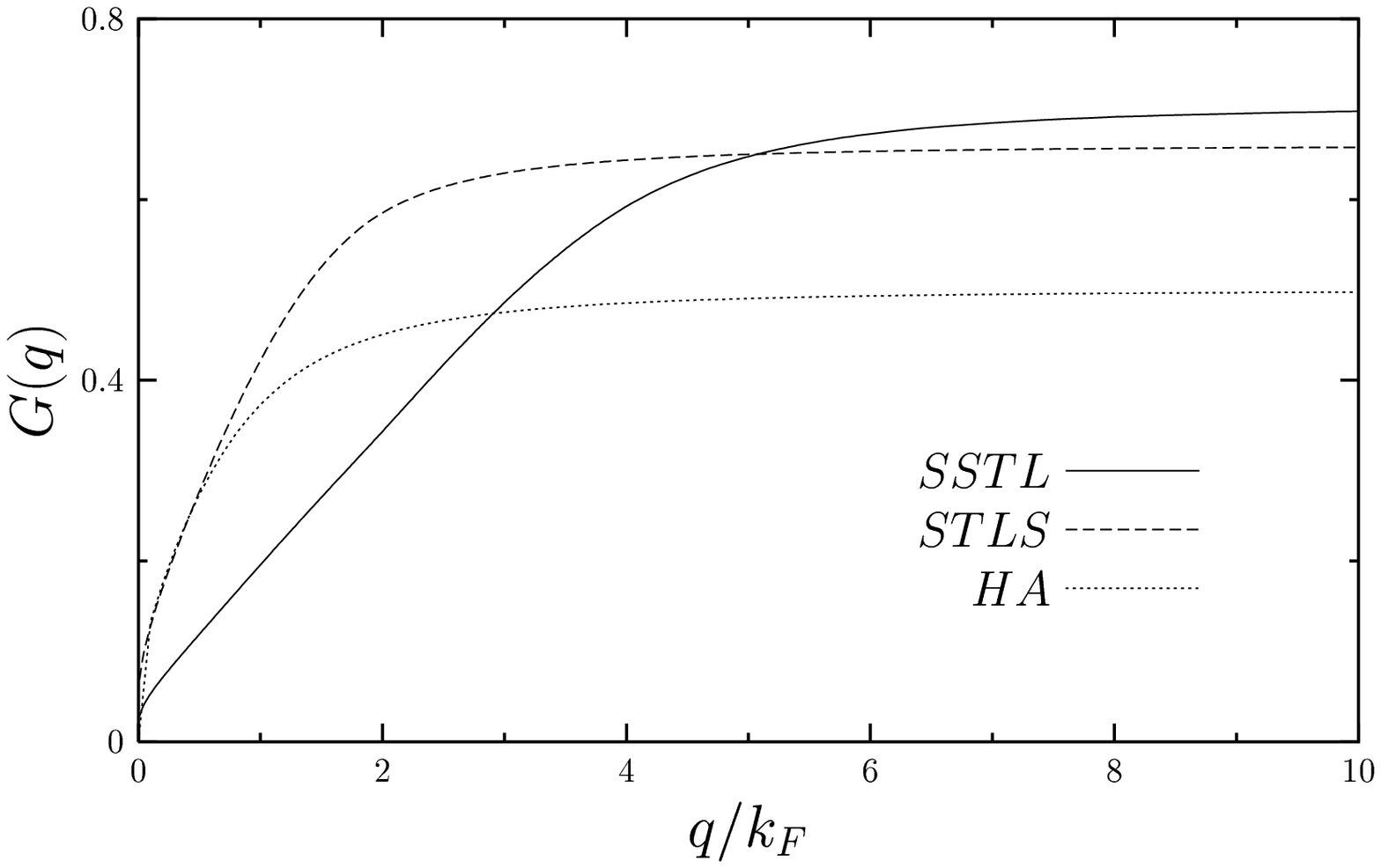}
\vskip 8.1cm
\caption{The density local field correction $G(q)$ in different
approximations for $r_s=1$ and ${\ds b=2a_B^{\star}}$.}
\end{figure}
\newpage

\begin{figure}
\vskip 1cm
    \includegraphics{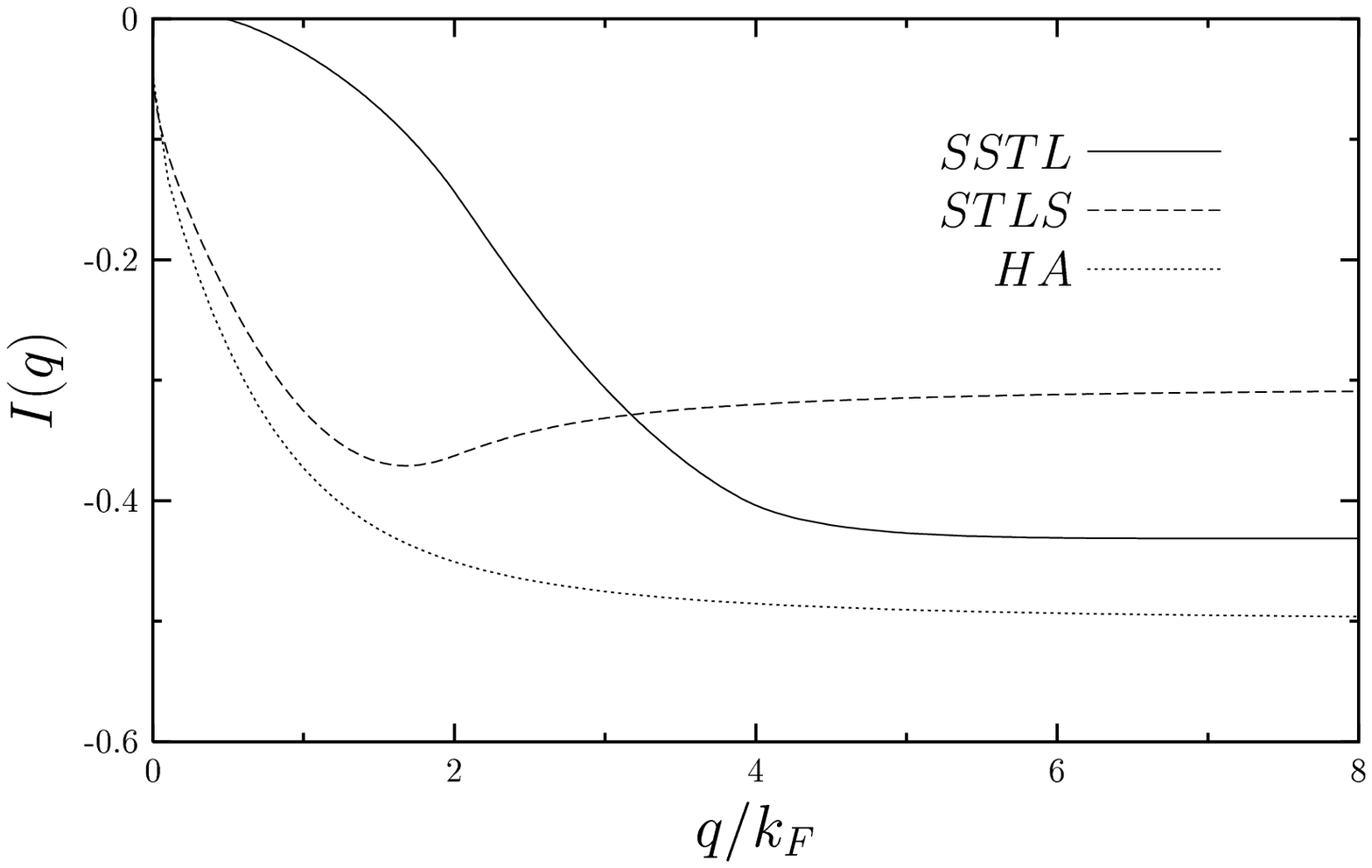}
\vskip 8.1cm
\caption{The spin--density local field correction $I(q)$ in different
approximations for $r_s=1$ and ${\ds b=2a_B^{\star}}$.}
\end{figure}

\begin{figure}
\vskip 1cm
    \includegraphics{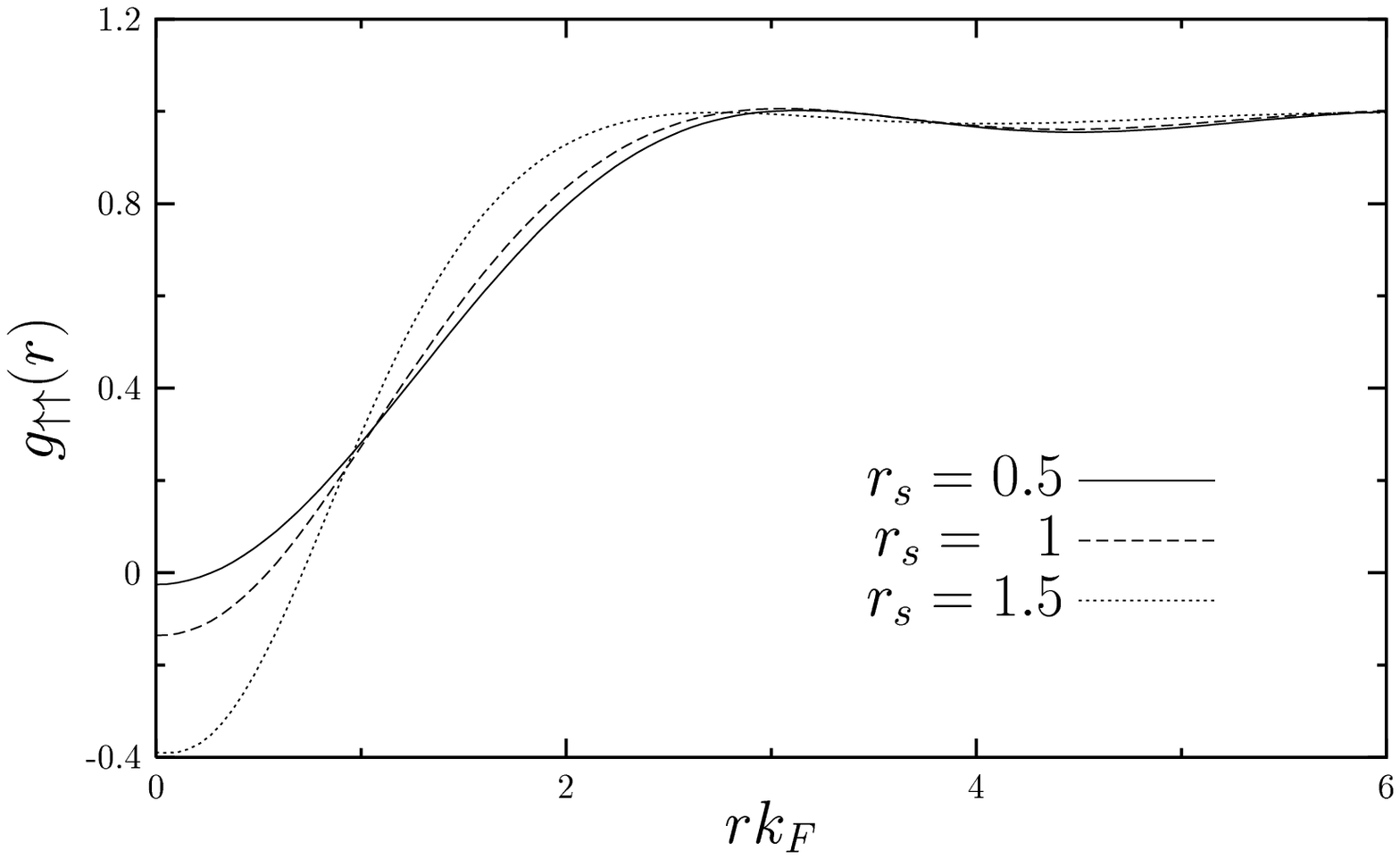}
\vskip 8.1cm
\caption{The spin dependent pair correlation function 
${\ds g_{\uparrow\uparrow}(r)}$ for ${\ds b=2a_B^{\star}}$ 
and several $r_s$ values in the SSTL approximation.}
\end{figure}
\newpage

\begin{figure}
\vskip 1cm
    \includegraphics{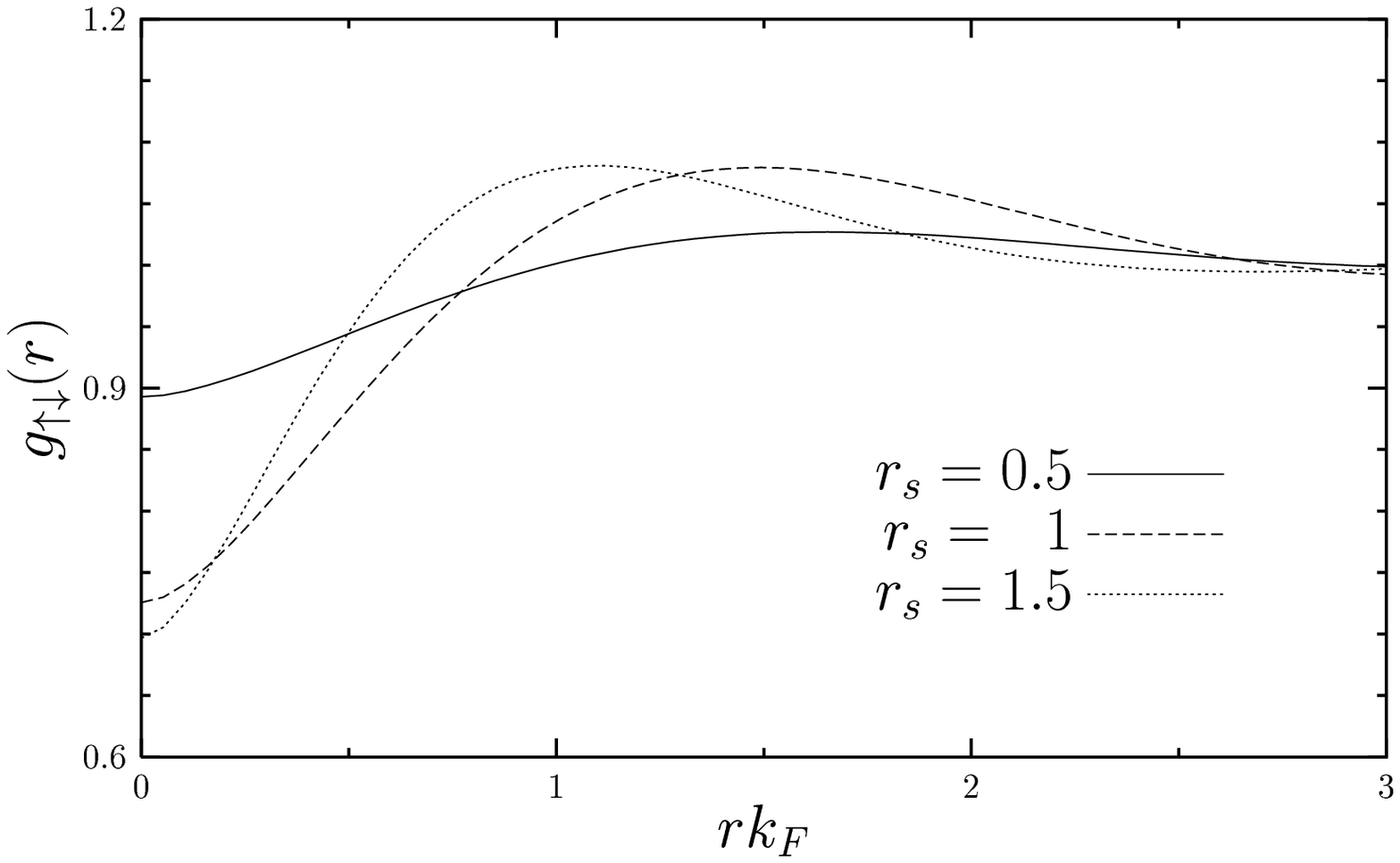}
\vskip 8.1cm
\caption{The spin dependent pair correlation function 
${\ds g_{\uparrow\downarrow}(r)}$ for ${\ds b=2a_B^{\star}}$ 
and several $r_s$ values in the SSTL approximation.}
\end{figure}

\begin{figure}
\vskip 1cm
    \includegraphics{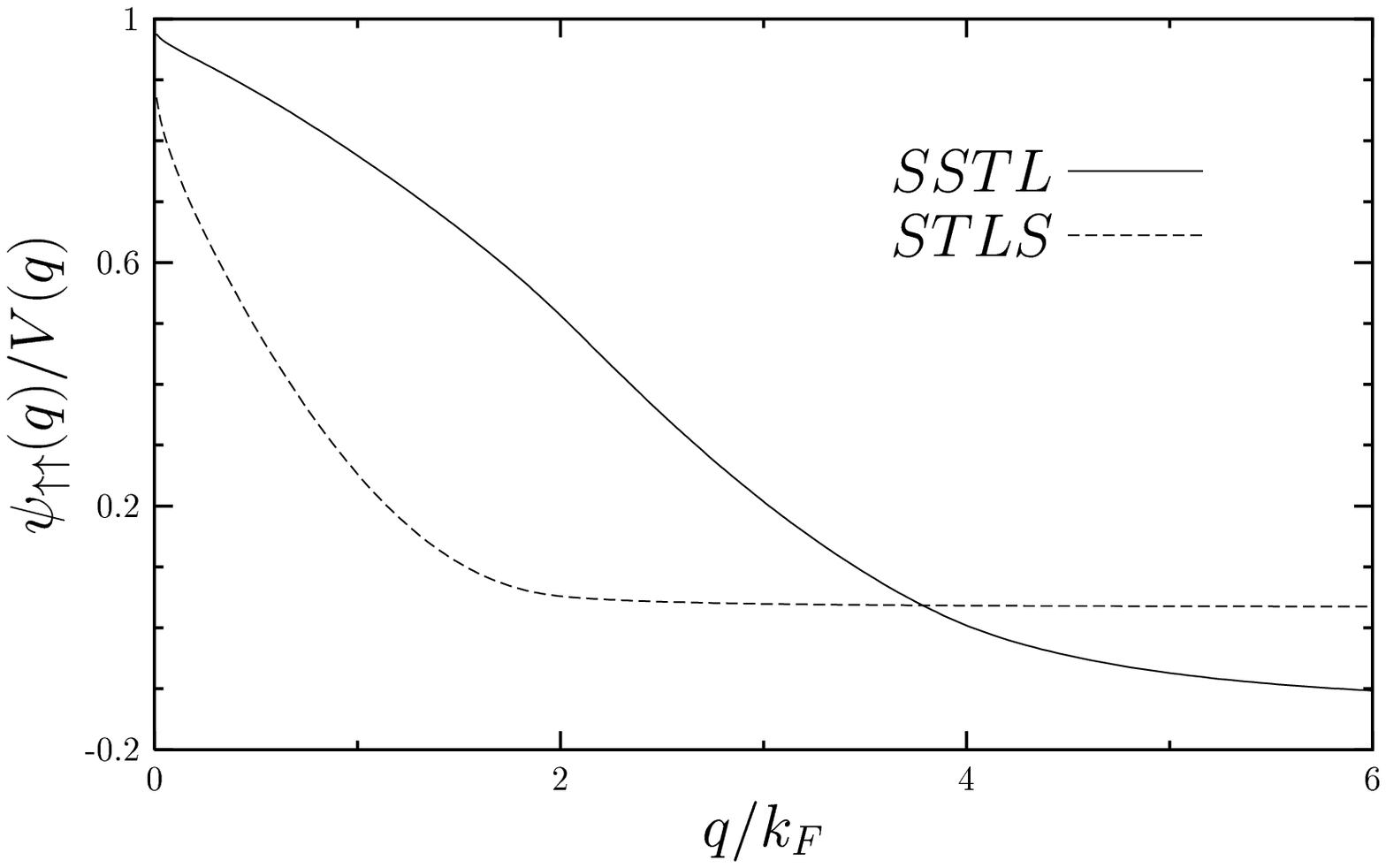}
\vskip 8.1cm
\caption{The spin dependent effective potential 
${\ds \psi_{\uparrow\uparrow}(q)}$ in units of $V(q)$ for $r_s=1$ 
and ${\ds b=2a_B^{\star}}$ in the SSTL and STLS approximations.}
\end{figure}
\newpage

\begin{figure}
\vskip 1cm
    \includegraphics{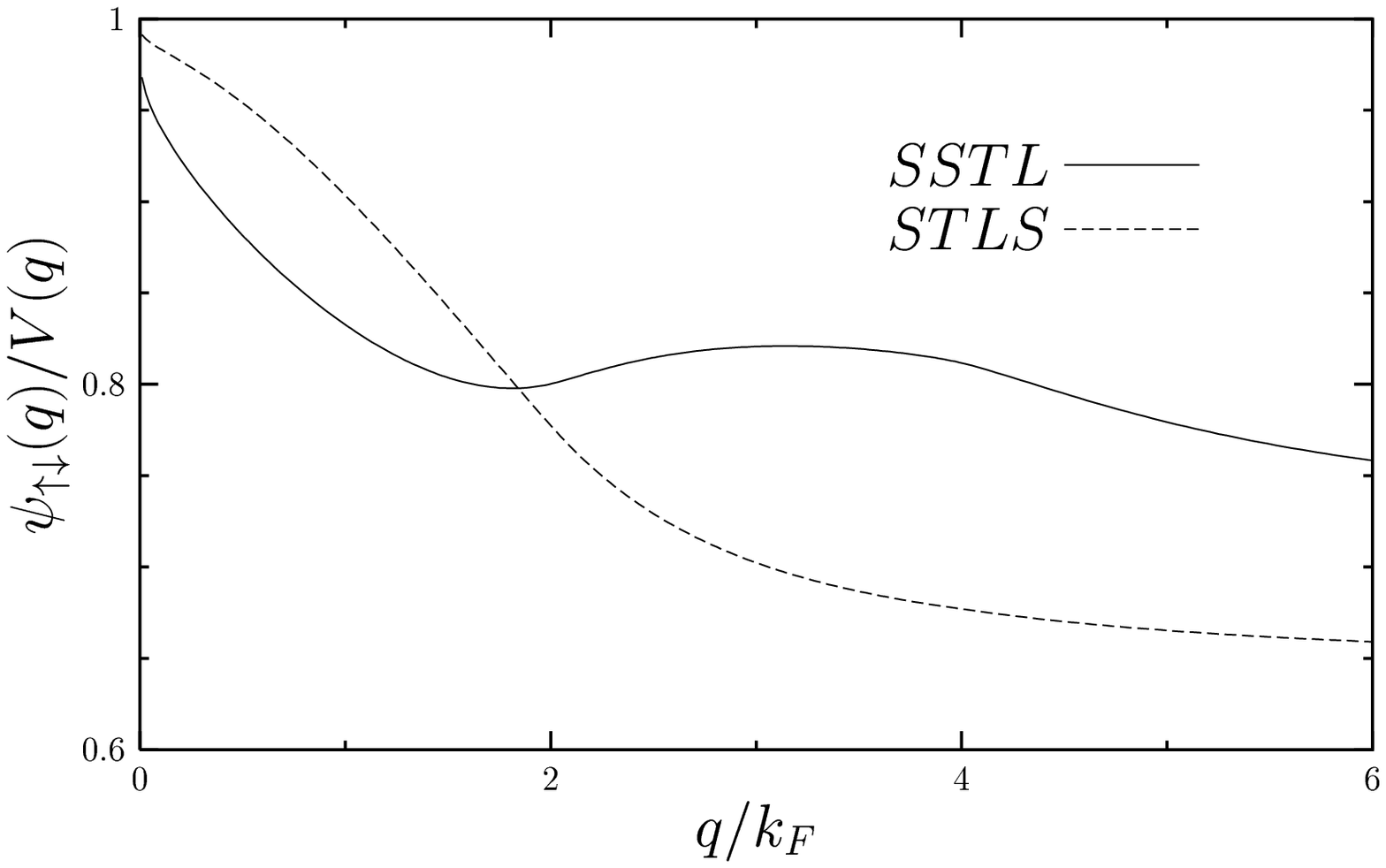}
\vskip 8.1cm
\caption{ The spin dependent effective potential 
${\ds \psi_{\uparrow\downarrow}(q)}$ in units of $V(q)$ for $r_s=1$ 
and ${\ds b=2a_B^{\star}}$ in the SSTL and STLS approximations.}
\end{figure}

\begin{figure}
\vskip 1cm
    \includegraphics{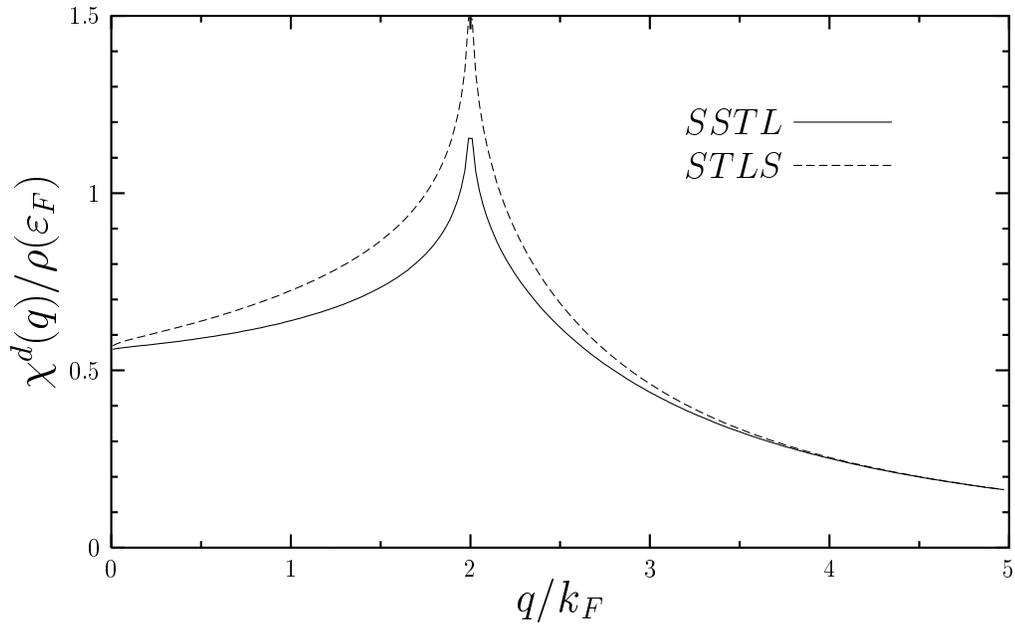}
\vskip 8.1cm
\caption{The static density response $\chi^d(q)$ in units of 
${\ds \rho(\varepsilon_F)}$ for $r_s=1$ and ${\ds b=2a_B^{\star}}$ 
in SSTL and STLS approximations.}
\end{figure}
\newpage

\begin{figure}
\vskip 1cm 
    \includegraphics{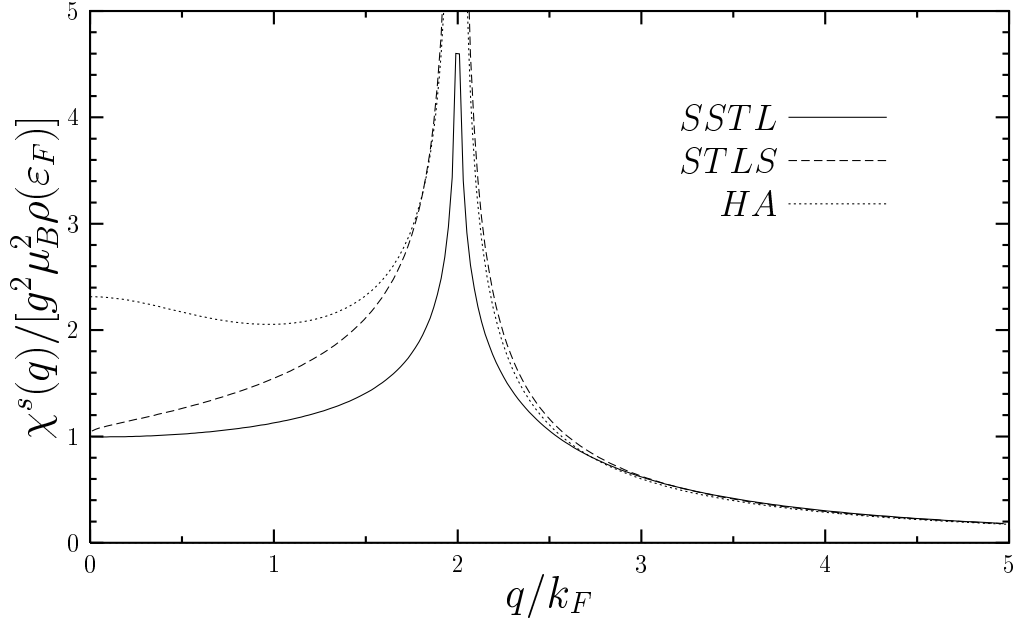}
\vskip 8.1cm 
\caption{The static spin response $\chi^s(q)$ in units of 
${g^2 \mu_B^2 \rho(\varepsilon_F)}$ for $r_s=1$ and 
${\ds b=2a_B^{\star}}$ in different approximations.}
\end{figure}

\begin{figure}
\vskip 1cm
    \includegraphics{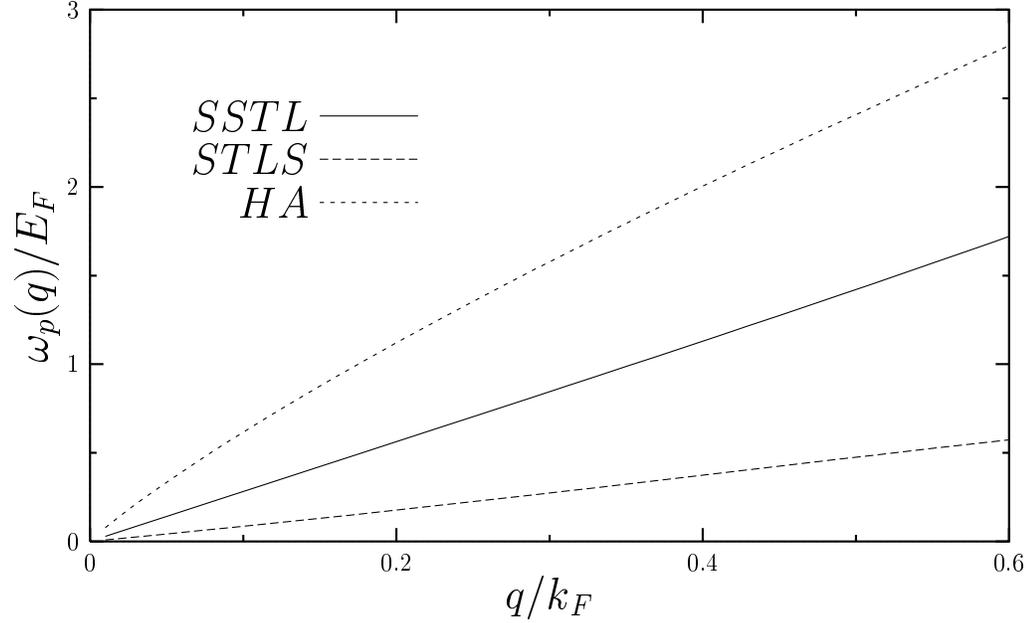}
\vskip 8.1cm
\caption{The dispersion of the paramagnon peak $\omega_p(q)$ in 
units of ${E_F}$ for $r_s=1.5$ and ${\ds b=2a_B^{\star}}$
in different approximations.}
\end{figure}
\end{document}